\documentclass[showpacs,preprint,amsmath,amssymb]{revtex4}
\usepackage{graphicx}
\usepackage{dcolumn}
\usepackage[usenames]{color}
\usepackage{bm}

\begin{document}

\title{Biaxial nematic phases in fluids of hard board-like particles}

\author{Y. Mart\'{\i}nez-Rat\'on}
\affiliation{Grupo Interdisciplinar de Sistemas Complejos (GISC),
Departamento de Matem\'{a}ticas,Escuela Polit\'{e}cnica Superior,
Universidad Carlos III de Madrid, Avenida de la Universidad 30, E--28911, 
Legan\'{e}s, Madrid, Spain}

\author{S. Varga}
\affiliation{Institute of Physics and Mechatronics, University of Pannonia, P.O. Box 158, Veszpr\'em H-8201, Hungary}

\author{E. Velasco}
\affiliation{Departamento de F\'{\i}sica Te\'orica de la Materia Condensada
and Instituto de Ciencia de Materiales Nicol\'as Cabrera,
Universidad Aut\'onoma de Madrid, E-28049 Madrid, Spain.}

\date{\today}

\begin{abstract}
We use density-functional theory, of the fundamental-measure type, to study the 
relative stability of the biaxial
nematic phase, with respect to non-uniform phases such as smectic and columnar,
in fluids made of hard board-like particles with sizes $\sigma_1>\sigma_2>\sigma_3$. 
A restricted-orientation (Zwanzig) approximation is adopted.
Varying the ratio $\kappa_1=\sigma_1/\sigma_2$ while keeping $\kappa_2=\sigma_2/\sigma_3$,
we predict phase diagrams for various values of $\kappa_2$ which include all the uniform phases:
isotropic, uniaxial rod- and plate-like nematics, and biaxial nematic. In addition,
spinodal instabilities of the uniform phases with respect to fluctuations of the
smectic, columnar and plastic-solid type, are obtained. In agreement with recent
experiments, we find that the biaxial nematic phase begins to be stable for
$\kappa_2\simeq 2.5$. Also, as predicted by previous theories and simulations
on biaxial hard particles, we obtain a region of biaxility centred on $\kappa_1\approx
\kappa_2$ which widens as $\kappa_2$ increases. For $\kappa_2\agt 5$ the region
$\kappa_2\approx\kappa_1$ of the packing-fraction vs. $\kappa_1$ phase diagrams
exhibits interesting topologies which change qualitatively with $\kappa_2$.
We have found that an increasing biaxial shape anisotropy favours the formation of the
biaxial nematic phase. Our study is the first to apply FMT theory to biaxial particles and,
therefore, it goes beyond the second-order virial approximation. Our prediction that the phase diagram 
must be asymmetric is a genuine result of the present approach, which is not accounted for by
previous studies based on second-order theories.

\end{abstract}

\pacs{Valid PACS appear here}

\maketitle

\definecolor{Red}{rgb}{1.00,0.00,0.00}

\section{Introduction}

The existence of biaxial nematic (N$_{\rm B}$) phases is a problem of great fundamental and 
practical interest, since these phases could be used in fast electro-optical devices \cite{Contro2}.
After their prediction by Freiser \cite{Freiser}, a great
effort has been devoted to their theoretical analysis \cite{Theo1,Theo2,Sim1,Sim2,Photinos1,Zannoni,Photinos2,VargaB} and 
experimental observation. The N$_{\rm B}$ phase was detected in a lyotropic
fluid \cite{Yu} and, more recently, in liquid crystals made of bent-core
organic molecules \cite{PRL1,PRL2,Galerne,Reply}; however, in the latter case controversial issues 
about the correct identification of the N$_{\rm B}$ phase still remain \cite{Contro1,Contro2,Galerne,Reply,Contro3}.
Biaxial phases have been predicted \cite{Alben} and observed in computer simulations \cite{Cuetos}
of mixtures of rod-like and plate-like particles. However, strong competition between biaxiality and
demixing is expected \cite{Stroobants,Roij,Varga,Wensink,MR}, 
and demixing could preempt biaxiality altogether.

Recently, a promising approach based on colloidal dispersions of mineral particles has been proposed.
Goethite particles with a board-like shape of sizes 
$\sigma_1>\sigma_2>\sigma_3$ and exhibiting short-ranged repulsive interactions 
were synthesised and observed to form colloidal liquid-crystalline phases
of nematic, smectic and columnar type \cite{Lemaire1,Lemaire2,Vroege1}.
Interestingly, a biaxial nematic phase has been observed \cite{Vroege2} when 
$\sigma_1/\sigma_2\approx\sigma_2/\sigma_3=3$. In this case
the N$_{\rm B}$ phase seems to be more stable than the competing smectic and columnar 
phases. Indeed, theory and simulation predicted that the biaxial nematic phase will be
formed when $\sigma_1/\sigma_2\approx\sigma_2/\sigma_3$ \cite{Theo1,Theo2,Sim2}.
 
The existence of biaxial phases in these colloidal particles opens a new avenue for theoretical
research. Their short-ranged repulsive interactions and well-defined shape allows for
the application of sophisticated density-functional theories for hard particles. 
In particular, these studies may help to understand why the biaxial phase is so hard
to stabilise and if competition with other more highly ordered phases, such as smectic and 
columnar, could be at work. In this connection, the early work by Somoza and Tarazona \cite{Somoza1,Somoza2},
who used a weighted-density functional theory applied to hard oblique cylinders, found stable smectic
phases and only a very limited region of biaxial-nematic stability. 
In more recent work, Taylor and Herzfeld \cite{Taylor} considered a fluid of hard spheroplatelets
(which are similar to board-shaped particles but with rounded corners)
and examined its phase behaviour using the scaled-particle theory of Cotter
\cite{Cotter}, which incorporates the excluded volume associated with two particles,
combined with a free-volume theory. The predicted phase diagram again showed that
biaxial-nematic stability was almost entirely suppressed by the smectic phase.
A recent simulation work \cite{Escobedo} analysed the full phase diagram of board-like
particles with square cross section, $\sigma_1=\sigma_2$ and $0.125<\sigma_3/\sigma_1<0.5$. The
first condition is not compatible with the formation of the the biaxial-nematic phase. To our knowledge, 
no simulation studies have been undertaken so far concerning more suitable size parameters.

Here we report on a density-functional theory where the phase stability of a fluid
of hard board-like particles is explored. Our aim is to examine the phase behaviour
of the goethite-based particles used in the experiments of van del Pol et al. \cite{Vroege2},
focusing on the formation of the N$_{\rm B}$ phase and their interplay with the uniaxial nematic
(N$_{\rm U}$) phases and the smectic (S) and columnar (C) phases. The approach is based on a
fundamental-measure theory (FMT) for hard parallelepipeds \cite{Yuri} in the so-called Zwanzig or
restricted-orientation approximation. Using this theory, we predict islands of stability for
the N$_{\rm B}$ as a function of the size parameters. These islands are not based on rigorous
boundaries, as the effect of the S and C phases is estimated only via bifurcation analyses
(not via costly free-energy minimisations).
However, our work should provide trends as to which particles sizes are more optimal to
stabilise the N$_{\rm B}$ phase. As a general rule, the particle aspect ratios should be
larger than those explored in the experiments. Our work extends and improves a previous
theoretical investigation \cite{Taylor} which considered very limited particle aspect ratios. 
Also our study, which is the first to adopt the FMT approach for biaxial particles and, in particular,
for hard board-like particles, may have some relevance in display applications based on fast-switching
particles that reorient by rotating their short axes instead of the long ones, something that can be
achieved with biaxial nematic phases. Finally, 
polydispersity in the sizes may play a key role when comparing our results with the experiments, but
was not taken into account in the present work.

In Section \ref{II} we define the particle model, and summarise the FMT theory and order parameters 
used to characterise uniaxial and biaxial order. Also, we write the bifurcation equations
solved to obtain the spinodal lines that define instability against non-uniform types
of ordering. Section \ref{III} presents the results, and in Section \ref{IV} we give some
conclusions.

\section{Model}
\label{II}

\begin{figure}[h]
\includegraphics[width=2.8in]{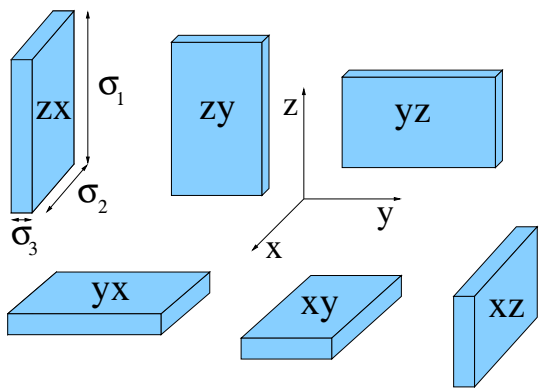}
\caption{\label{fig1}Schematic of board-like particle with definition of the three
sizes: $\sigma_1$ (length), $\sigma_2$ (width) and $\sigma_3$ (thickness). 
A depiction of the six possible particle orientations is included. Labels $\mu\nu$ (with $\{\mu,\nu\}=
\{x,y,z\}$ and $\mu\ne\nu$) indicate orientation with longest axis parallel to 
$\mu$ and second axis along $\nu$.}
\end{figure}

The system consists of a collection of biaxial hard parallelepipeds with 
edge lengths $\sigma_i$ ($i=1,2,3$) satisfying $\sigma_1>\sigma_2>\sigma_3$
(Fig. \ref{fig1}). We define the size ratios
\begin{eqnarray}
\kappa_1=\frac{\sigma_1}{\sigma_2},\hspace{0.8cm}
\kappa_2=\frac{\sigma_2}{\sigma_3},
\end{eqnarray}
as the ratio of length-to-width and width-to-thickness, respectively.
%
%
Now the fluid of identical particles is mapped onto an equivalent fluid
mixture of six species as follows. Within the Zwanzig approximation, where particle
orientations are restricted to the three orthogonal Cartesian axes,
particles oriented in the same way are considered to belong to the
same species, and six unequivalent species result. Then, the density profiles 
$\rho_{\mu\nu}({\bf r})$ will be the fundamental quantities that characterise the 
equilibrium properties of the fluid; the subindices
$\mu$ and $\nu$ represent the directions parallel to the first (length) 
and second (width) main axes of the particles, respectively.

According to FMT \cite{Yuri}, 
the excess part of the free-energy density in thermal-energy units $kT\equiv\beta^{-1}$ 
is given by
\begin{eqnarray}
\Phi=-n_0\ln(1-n_3)+\frac{{\bf n}_1\cdot {\bf n}_2}{1-n_3}
+\frac{n_{2x}n_{2y}n_{2z}}{(1-n_3)^2},
\end{eqnarray}
where the weighted densities $n_{\alpha}({\bf r})$ are defined through 
convolutions:
\begin{eqnarray}
n_{\alpha}({\bf r})=\sum_{\mu\nu}\int d{\bf r}'\rho_{\mu\nu}({\bf r}')
\omega_{\mu\nu}^{(\alpha)}({\bf r}-{\bf r}'),
\end{eqnarray}
with $\alpha=0,1,2$ or $3$.
The weights $\omega_{\mu\nu}^{(\alpha)}({\bf r})$, defined in \cite{Yuri} for uniaxial 
parallelepipeds, are extended in the present work to the general case of biaxial particles as
\begin{eqnarray}
\omega^{(0)}_{\mu\nu}({\bf r})&=&\frac{1}{8}\prod_{i=1}^3 \delta_{\mu\nu}(x_i),\quad 
\omega^{(3)}_{\mu\nu}({\bf r})=\prod_{i=1}^3 \theta_{\mu\nu}(x_i), \\
\boldsymbol{\omega}^{(1)}_{\mu\nu}({\bf r})&=&\frac{1}{4}\left(
\theta_{\mu\nu}(x)\delta_{\mu\nu}(y)\delta_{\mu\nu}(z), 
\delta_{\mu\nu}(x)\theta_{\mu\nu}(y)\delta_{\mu\nu}(z),\right.\nonumber\\
&&\left.\delta_{\mu\nu}(x)\delta_{\mu\nu}(y)\theta_{\mu\nu}(z)\right),\\
\boldsymbol{\omega}^{(2)}_{\mu\nu}({\bf r})&=&\frac{1}{2}\left(
\delta_{\mu\nu}(x)\theta_{\mu\nu}(y)\theta_{\mu\nu}(z), 
\theta_{\mu\nu}(x)\delta_{\mu\nu}(y)\theta_{\mu\nu}(z),\right.\nonumber\\
&&\left.\theta_{\mu\nu}(x)\theta_{\mu\nu}(y)\delta_{\mu\nu}(z)\right),
\end{eqnarray}
where we have defined 
\begin{eqnarray}
\delta_{\mu\nu}(x_i)&=&\delta\left(\frac{\sigma_{\mu\nu}^i}{2}-|x_i|\right),\\ 
\theta_{\mu\nu}(x_i)&=&\Theta\left(\frac{\sigma_{\mu\nu}^i}{2}-|x_i|\right),
\end{eqnarray}
with $\delta(x)$ and $\Theta(x)$ the Dirac-delta and Heviside functions respectively, while 
$\sigma_{\mu\nu}^i=\sigma_3+(\sigma_1-\sigma_3)\delta_{i\mu}+(\sigma_2-\sigma_3)\delta_{i\nu}$ 
denotes the length of the biaxial particle with orientations $\mu$ and $\nu$ along the $i$-th direction. 
We have used the notation $x_1=x$, $x_2=y$ and $x_3=z$.

The total free-energy density is obtained by adding the ideal free-energy contribution,
\begin{eqnarray}
\beta {\cal F}&=&\int_V d{\bf r}\left[\Phi_{\rm{id}}({\bf r})+\Phi({\bf r})\right],\\\
\Phi_{\rm{id}}({\bf r})&=&\sum_{\mu\nu}\rho_{\mu\nu}({\bf r})
\left(\ln{\rho_{\mu\nu}}({\bf r})-1\right),
\end{eqnarray}
where $V$ is the fluid volume.
The equilibrium state of the fluid follows by minimisation of $\beta{\cal F}/V$ with respect
to the six independent densities $\rho_{\mu\nu}({\bf r})$ at fixed total density per unit cell 
(the latter being one-, two- or three-dimensional for the smectic, columnar or crystalline phases,
respectively).

\subsection{Uniform phases: isotropic, uniaxial and biaxial nematics}

In the uniform phase the free-energy densities can be written as
\begin{eqnarray}
\Phi_{\rm{id}}&=&\rho\left(\ln \rho-1+\sum_{\mu\nu}x_{\mu\nu}\ln x_{\mu\nu}\right)\\\nonumber\\
\Phi&=&-\xi_0\ln(1-\xi_3)+\frac{\boldsymbol{\xi}_1\cdot\boldsymbol{\xi}_2}{1-\xi_3}+
\frac{\xi_{2x}\xi_{2y}\xi_{2z}}{(1-\xi_3)^2},
\end{eqnarray}
with $\xi_{\alpha}=\rho\sum_{\mu\nu}x_{\mu\nu}w^{(\alpha)}_{\mu\nu}$. The total density and packing fraction
are $\xi_0=\rho$ and $\xi_3=\eta$, respectively, while $x_{\mu\nu}$ are the
molar fractions. The coefficients $w_{\mu\nu}^{(\alpha)}$ are defined as 
$w_{\mu\nu}^{(\alpha)}=\int d{\bf r}\omega^{(\alpha)}_{\mu\nu}({\bf r})$, with 
$w_{\mu\nu}^{(0)}=1$, $w_{\mu\nu}^{(3)}=\sigma_1\sigma_2\sigma_3$, and
\begin{eqnarray}
{\bf w}_{\mu\nu}^{(1)}&=&\left(\sigma_{\mu\nu}^{x},\sigma_{\mu\nu}^{y},\sigma_{\mu\nu}^{z}\right),\\
{\bf w}_{\mu\nu}^{(2)}&=&\left(\sigma_{\mu\nu}^{y}\sigma_{\mu\nu}^{z},\sigma_{\mu\nu}^{z}
\sigma_{\mu\nu}^{x},\sigma_{\mu\nu}^{x}\sigma_{\mu\nu}^{y}\right).
\end{eqnarray}
The fluid pressure can be calculated from
\begin{eqnarray}
\beta p=\frac{\xi_0}{1-\xi_3}+\frac{\boldsymbol{\xi}_1\cdot\boldsymbol{\xi}_2}{(1-\xi_3)^2}+
\frac{2\xi_{2x}\xi_{2y}\xi_{2z}}{(1-\xi_3)^3},
\end{eqnarray}
while the chemical potential is
\begin{eqnarray}
\beta\mu=\rho^{-1}\left(\beta p+\Phi_{\rm{id}}+\Phi\right).
\end{eqnarray}
To find the equilibrium uniform phases we fix the packing fraction $\eta$ and 
minimize the total free-energy density 
$\Phi_{\rm{id}}+\Phi$ with respect to the molar fractions $\{x_{\mu\nu}\}$, with the 
obvious constraint $\sum_{\mu\nu}x_{\mu\nu}=1$. 

Let us now discuss how the orientational order can be defined quantitatively. In the present model, the
usual order parameter tensor 
\begin{eqnarray}
Q_{\alpha\beta}=\frac{1}{2}\left(3\langle u_{\alpha}^{j}u_{\beta}^{j}\rangle-\delta_{\alpha\beta}\right),
\end{eqnarray}
(with $u_{\alpha}^{j}$ the projection of the unit vector of the long axis of particle $j$
on the $\alpha$ Cartesian axis, the average taken over all particles) becomes diagonal,
\begin{eqnarray}
Q_{\alpha\beta}&=&\frac{1}{2}\left(3\sum_{\mu\nu} x_{\mu\nu}\delta_{\mu\alpha}\delta_{\mu\beta}-\delta_{\alpha\beta}\right)
\nonumber\\\nonumber\\&=&\frac{1}{2}\left(3\sum_{\nu\neq\alpha}x_{\alpha\nu}-1\right)\delta_{\alpha\beta}.
\end{eqnarray}
Assuming that the first nematic director is parallel to the $z$ axis while the second lies along the $x$ axis, our uniaxial 
and biaxial order parameters are, respectively 
\begin{eqnarray}
&&Q\equiv Q_{zz}=\frac{1}{2}\left(3\sum_{\nu\neq z}x_{z\nu}-1\right),
\nonumber\\\nonumber\\
&&\delta\equiv Q_{xx}-Q_{yy}=\frac{3}{2}\left(\sum_{\nu\neq x}x_{x\nu}-\sum_{\nu\neq y}x_{y\nu}\right).
\end{eqnarray}
The order parameter $\delta$ measures the difference in the fraction of particles pointing along the $x$ and $y$ 
axes. If there are no particles in the $xy$ plane, then $\delta=0$.
Therefore, the order parameter matrix has the usual diagonal form
\begin{eqnarray}
\hat{Q}=
\begin{pmatrix}
-\displaystyle{\frac{Q-\delta}{2}} & 0 & 0\\
0 & -\displaystyle{\frac{Q+\delta}{2}} & 0\\
0 & 0 & Q
\end{pmatrix}.
\end{eqnarray}
However, the biaxial order parameter $\delta$, which is zero for perfect unaxial order and non-zero when there is some
amount of biaxiality, has the problem that, for perfect biaxial alignment, $\delta=0$. A better tensor quantity to characterize 
biaxiality involves the other two particles axes:
\begin{eqnarray}
\Delta_{\alpha\beta}=\frac{1}{2}\left(\langle e_{\alpha}^{j}e_{\beta}^{j}\rangle 
-\langle s_{\alpha}^{j}s_{\beta}^{j}\rangle\right),
\end{eqnarray}
where $e^{j}_{\alpha}$ and $s^{j}_{\alpha}$ are the projections of the second and third axes of the $j$-th particle on the
$\alpha$ Cartesian axis, respectively. For our system we again find a diagonal tensor,
\begin{eqnarray}
\Delta_{\alpha\beta}=\frac{1}{2}\left(\sum_{\mu\neq \alpha}x_{\mu\alpha}
-\sum_{\mu\neq\nu,\nu\neq \alpha}x_{\mu\nu}\right)\delta_{\alpha\beta}.
\end{eqnarray}
These magnitudes can be writen as a function of new order parameters $S$ and $\Delta$ as 
\begin{eqnarray}
\Delta_{xx}=-\frac{(S-\Delta)}{2},\quad \Delta_{yy}=-\frac{(S+\Delta)}{2},\quad 
\Delta_{zz}=S.
\end{eqnarray}
In terms of the mole fractions $x_{\mu\nu}$, these order parameters can be explicitly written as
\begin{eqnarray}
S&=&\frac{1}{2}\left(x_{xz}-x_{xy}+x_{yz}-x_{yx}\right),\\
\Delta&=&\frac{1}{2}\left[x_{yx}-x_{xy}+x_{xz}-x_{yz}+2(x_{zx}-x_{zy})\right].
\end{eqnarray} 
Note that the first and second nematic directors have been chosen to be parallel to 
the $z$ and $x$ axes, respectively. While $S$ measures the extent of biaxial ordering along the $x$ and $y$ axes, 
$\Delta$ is the global biaxial parameter which 
is zero for the uniaxial nematic phase (since uniaxial nematic symmetry 
implies $x_{yx}=x_{xy}$, $x_{xz}=x_{yz}$ and $x_{zx}=x_{zy}$), 
and unity for perfect biaxial ordering (for which $x_{\mu\nu}=0,\forall \mu,\nu\neq z,x$, while 
$x_{zx}=1$).

\subsection{Non-uniform phases}

The existence of spatially non-uniform phases, such as the smectic or columnar phases,
has been assessed via bifurcation analysis from a uniform (either isotropic or nematic) phase.
The relevant response function of the system is the direct correlation function,
$c_{\mu\nu,\tau\theta}({\bf r}-{\bf r}')$.
We define the $6\times 6$ matrix ${\cal S}$ with elements 
\begin{eqnarray}
{\cal S}_{\alpha\beta}(\eta,{\bf q})=\delta_{\alpha\beta}
-\rho_{\mu\nu}\hat{c}_{\mu\nu,\tau\theta}({\bf q}),
\end{eqnarray}
where $\alpha=\mu+\nu-2$ if $\mu<\nu$ and $\alpha=\mu+\nu+1$ if $\mu>\nu$ while
$\beta=\tau+\theta-2$ if $\tau<\theta$ and $\beta=\tau+\theta+1$ if 
$\tau>\theta$. The Fourier transforms of the direct correlation functions, 
\begin{eqnarray}
-\hat{c}_{\mu\nu,\tau\theta}({\bf q})=\sum_{\alpha\beta}\frac{\partial^2\Phi}{\partial\xi_{\alpha}
\partial\xi_{\beta}}\hat{\omega}^{(\alpha)}_{\mu\nu}({\bf q})
\hat{\omega}^{(\beta)}_{\tau\theta}({\bf q}),
\end{eqnarray}
with ${\bf q}$ the wave vector,
can be calculated from the Fourier transforms $\hat{\omega}^{(\alpha)}_{\mu\nu}({\bf q})$ of the 
correspondings weights $\omega^{(\alpha)}_{\mu\nu}({\bf r})$.
To calculate the spinodal or 
bifurcation curves, we need to solve the equation
\begin{eqnarray}
D_s(\eta,{\bf q})=0,\quad \boldsymbol{\nabla} D_s(\eta,{\bf q})=0,
\end{eqnarray}
where $D_s=\text{det}\left({\cal S}\right)$. From here we find, by iteration, the values of 
packing fraction $\eta$ and wave vector ${\bf q}$ at the spinodal instability. Note that, 
during the iterative process, for each value of $\eta$ one needs to find the molar fractions
$x_{\mu\nu}$ via minimization of the free-energy density.

\section{Results}
\label{III}

\begin{figure}[h]
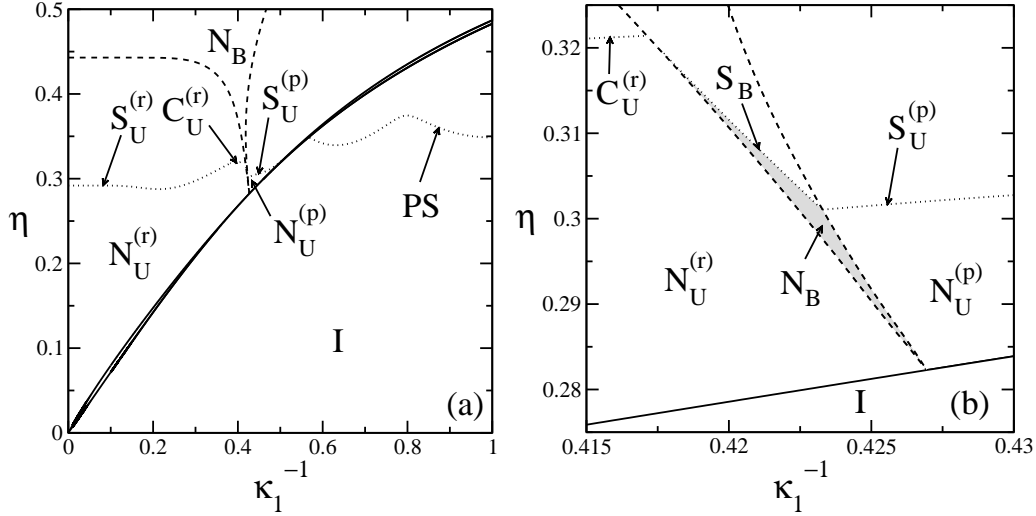

\includegraphics[width=2.6in]{Fig2a.eps}
\includegraphics[width=2.75in]{Fig2b.eps}
\caption{(a) Phase diagram in the plane $\eta-\kappa_1^{-1}$ for hard parallelepipeds with 
$\kappa_2=2.5$. (b) A zoom of the region about $\kappa_1^{-1}\approx 0.4$. Labels indicate regions of
stability of the different phases, or type of ordering with respect to which bifurcation curves are
computed (see text for key to labels). Solid and dashed curves indicate first and second order phase 
transition, respectively. Dotted curves are bifurcation lines to the corresponding 
non-uniform phase from the stable bulk uniform phases. In (b) the shaded region corresponds to the 
estimated maximum region of stability of the N$_{\rm B}$ phase.}
\label{fig2}
\end{figure}

We have applied our FMT approach to study a fluid of board-shaped particles with 
fixed values of the width-to-thickness ratio, i.e. $\kappa_2$. Phase diagrams 
in the plane $\eta$--$\kappa_1^{-1}$ were produced in each case.

Our first result is Fig. \ref{fig2}(a), which depicts the phase diagram for the case $\kappa_2=2.5$. 
We identify one isotropic phase (I), with $Q=\Delta=0$, and two uniaxial nematic
phases, N$_{\rm U}^{\rm (r)}$ and N$_{\rm U}^{\rm (p)}$, with $Q\neq 0$ and $\Delta=0$. 
These two phases can be distinguished 
by the direction the uniaxial director takes with respect to the sides of the
particles. In N$_{\rm U}^{\rm (r)}$, the director points along the long particle side (`rod-like' 
nematic), while in N$_{\rm U}^{\rm (p)}$ the director lies along the short side (`plate-like'
nematic). In between these two phases a region of biaxial nematic, N$_{\rm B}$, appears
as a wedge centred at $\kappa_1\approx\kappa_2$, as predicted previously \cite{Theo1,Theo2,Sim2}. 
This region opens up as density is increased. Concerning the I-N$_{\rm U}$ transitions, they are always 
of first order, except at some particular point in the $\eta-\kappa_1$ plane, where the density gap 
is exactly zero.
This is the point where the continuous N$_{\rm U}^{\rm (r)}$-N$_{\rm B}$ and N$_{\rm B}$-N$_{\rm U}^{\rm (p)}$ 
transition lines meet (Landau point). These transitions are always continuous, regardless of the value of $\kappa_1$.
The topology of the phase diagram, as far as isotropic and nematic phases are concerned, is
similar to that obtained previously by Camp and Allen \cite{Sim2} for hard biaxial ellipsoids.
However, it is slightly different from that of Taylor and Herzfeld \cite{Taylor} for hard
spheroplatelets since, in that case, the N$_{\rm B}$ region does not extend down to the
I-N$_{\rm U}$ transition boundary. The restricted-orientation approximation was invoked in \cite{Taylor}
to explain the discrepancy with simulation. However, the fact that we are also using this
approximation points to a different explanation. The different particle geometries (a spheroplatelet is
generated by a sphere whose centre is constrained to a rectangle, and therefore has rounded
instead of sharp corners) may be one possibility.

One interesting feature of the N$_{\rm U}^{\rm (r)}$-N$_{\rm B}$ transition line is that it tends to
a constant packing fraction as $\kappa_1\to\infty$. As one approaches this limit, the value of the uniaxial
order parameter $Q$ tends to unity, and the transition exclusively involves a two-dimensional
ordering of the intermediate (secondary axis) particle side. Also, in the neighbourhood of the
value $\kappa_1=\kappa_2$, the biaxial phase becomes reentrant, as can be inferred from the negative
slope of the N$_{\rm B}$-N$_{\rm U}^{\rm (p)}$, visible in the zoom presented in Fig. \ref{fig2}(b), which later 
becomes positive. 

Overall the phase diagram is very asymmetric with respect to the condition $\kappa_1=\kappa_2$, in contrast with
models based on the second virial coefficient. 
This coefficient has 
rod-plate symmetry, and it is the inclusion (implicit in the present model) of third and higher virial coefficients 
that breaks the symmetry, producing an asymmetrical phase diagram. 

\begin{figure}[h]
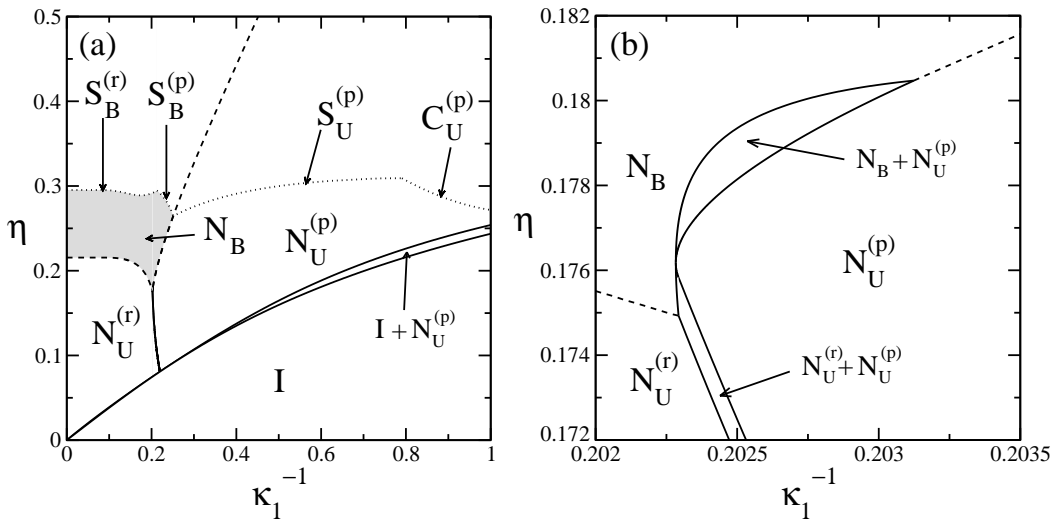

\includegraphics[width=2.6in]{Fig3a.eps}
\includegraphics[width=2.85in]{Fig3b.eps}
\caption{(a) Phase diagram in the plane $\eta-\kappa_1$
for hard biaxial parallelepipeds with $\kappa_2=5$. (b) A zoom of the
region about $\kappa_1^{-1}\approx 0.2$. Labels as in Fig. \ref{fig2}.}
\label{fig3}
\end{figure}

Let us turn to the spinodal lines. These are associated with instabilities of the stable bulk phase with respect to 
spatially non-uniform types of ordering. In Fig. \ref{fig2}, different bifurcation curves have been represented, 
corresponding to PS (plastic solid, with three-dimensional spatial order but no orientational order), S$_{\rm U}^{\rm (r)}$
and S$_{\rm U}^{\rm (p)}$ (uniaxial rod- and plate-like smectic order, respectively), S$_{\rm B}$ (biaxial
smectic order) and C$_{\rm U}^{\rm (r)}$ (uniaxial columnar order with the long particle axes parallel to the column axes).
In the region $\kappa_1\agt 1$ the I fluid is unstable with respect to the PS phase. This spinodal curve
crosses the I-N$_{\rm U}^{\rm (p)}$ coexistence region and connects to the S$_{\rm U}^{\rm (p)}$ spinodal curve,
which finally joins the stability region of the N$_{\rm B}$ phase. The region where the N$_{\rm U}^{\rm (p)}$
phase is stable is reduced to a very small triangular region. Inside the N$_{\rm B}$ region, we found a
short bifurcation line associated with S$_{\rm B}$ ordering [see enlarged area in Fig. \ref{fig2}(b)].
This line touches the left boundary of the N$_{\rm B}$ region, and smoothly continues into the uniaxial region
as a bifurcation line associated with columnar-type fluctuations, C$_{\rm U}^{\rm (r)}$ in Fig. \ref{fig2} \cite{C}, which in turn
is connected to the S$_{\rm U}^{\rm (r)}$ bifurcation curve. These calculations indicate that, for the
case $\kappa_2=2.5$, the region in which the N$_{\rm B}$ phase is stable is exceedingly small.
This coincides with the conclusion drawn for spheroplatelets by Taylor and Herzfeld \cite{Taylor}, who 
used a version of free-volume theory to study absolute stability of smectic, columnar and crystal phases. 
They concluded that the biaxial nematic phase is almost completely preempted by the uniaxial
and biaxial smectic phases.

The case $\kappa_2=2.5$ was chosen to be close to that considered by Taylor and Herzfeld \cite{Taylor}.
If $a$ is the radius of the sphere and $c\times b$ are the sizes of the rectangle that define the
spheroplatelet, with $c>b>a$, then the true sizes of the particle used in Ref. \cite{Taylor} are
$(c+2a)\times(b+2a)\times 2a$, and we can define length-to-width and width-to-thickness ratios
for this particle as $\kappa_1=(1+c/2a)/(1+b/2a)$ and $\kappa_2=1+b/2a$. In their work, 
Taylor and Herzfeld fixed the ratio $c/2a=5$ and varied $b/2a$. This is equivalent to fixing the
product $\kappa_1\kappa_2=6$. The condition $\kappa_1=\kappa_2$ is fulfilled for $\kappa_2=\sqrt{6}=
2.449...$, which is indeed where Taylor and Herzfeld find the N$_{\rm B}$ phase. Our choice for
$\kappa_2$ implies that our particles have the same side ratios close to the Landau point and therefore
that the phase-diagram topology should be almost identical in that region. This is true, save the
discrepancies mentioned above. Outside of this region we are exploring particle sizes differently.
In fact, in the following we go beyond Taylor and Herzfeld's work and consider other values of $\kappa_2$; 
we will see that changes in the phase diagram are very drastic and not just quantitative.
 
\begin{figure}[h]
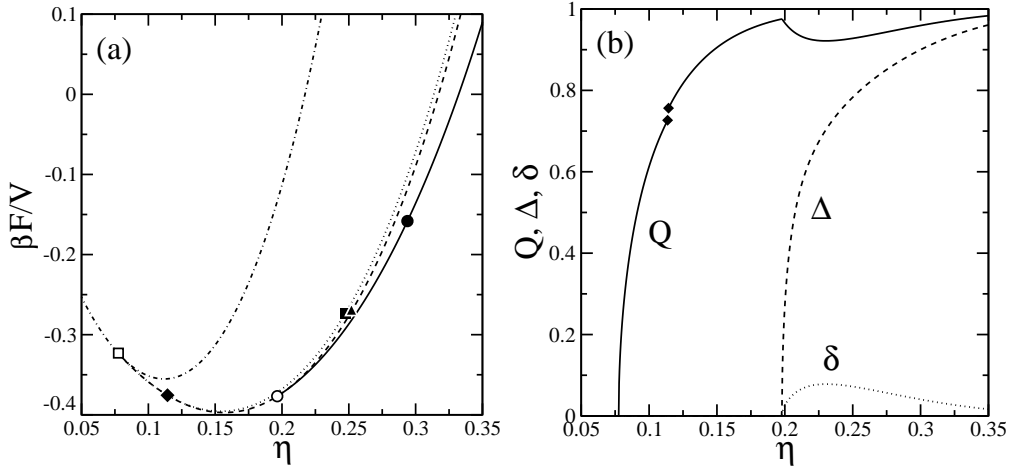

\includegraphics[width=2.6in]{Fig5a.eps}
\includegraphics[width=2.6in]{Fig5b.eps}
\caption{(a) Branches of the free energy density per unit thermal energy, $\beta{\cal F}/V$, as a function of packing
fraction $\eta$ for the case $\kappa_2=5$, $\kappa_1=4.76$ ($\kappa_1^{-1}=0.21$) and for the different phases.
Continuous curve: N$_{\rm B}$. Dashed curve: N$_{\rm U}^{\rm (p)}$. 
Dotted curve: N$_{\rm U}^{\rm (r)}$. Dash-dotted curve: I. The symbols indicate 
bifurcation points, as follows: open square, I-N$_{\rm U}^{\rm (p)}$; 
rhombus, N$_{\rm U}^{\rm (p)}$-N$_{\rm U}^{\rm (r)}$ 
(note that the last two transitions are of first order but very weak);
open circle: N$_{\rm U}^{\rm (r)}$-N$_{\rm B}$; filled circle: N$_{\rm B}$-S$_{\rm B}^{\rm (p)}$;
filled triangle and square: bifurcation from (unstable) N$_{\rm U}$ phases to a spatially
nonuniform phase. (b) Orientational order parameters as a function of packing fraction for
the same particle size parameters as in (a). The symbols indicate the discontinuity values of $Q$ caused by the
first-order nature of the N$_{\rm U}^{\rm (p)}$-N$_{\rm U}^{\rm (r)}$ transition.}
\label{frees}
\end{figure}

We have obtained the phase diagram for the case $\kappa_2=5$, which is plotted in Fig. \ref{fig3}(a).
New features are apparent. First, the N$_{\rm B}$ phase again appears for $\kappa_1\simeq\kappa_2$,
but now the topology of the phase diagram in this region changes dramatically [a zoom is provided
in Fig. \ref{fig3}(b)]. Now the N$_{\rm B}$ becomes disconnected from the I-N$_{\rm U}$ transition, and
there appears a direct, first-order transition between the two unaxial nematics, 
N$_{\rm U}^{\rm (r)}$ and N$_{\rm U}^{\rm (p)}$. This remarkable result, which has never been observed before
in the context of molecular theories for one-component hard anisotropic particles, 
resembles the phenomenology obtained by Taylor 
and Herzfeld and discussed above, but with the important difference that their reported N$_{\rm U}^{\rm (r)}$-N$_{\rm U}^{\rm (p)}$ 
transition is continuous (it is likely that, for larger values of $\kappa_2$, their theory would have predicted 
a first-order transition as well). The change in topology of the present case, with respect to the one obtained
for the case $\kappa_2=2.5$, can be explained using a Landau expansion for the free energy up to the fourth order in 
the order parameter tensor $Q_{\alpha\beta}$.
Imposing particular conditions on the expansion coefficients, the continuous N$_{\rm B}$-N$_{\rm U}^{\rm (r,p)}$ 
transitions can be replaced by a first-order N$_{\rm U}^{\rm (r)}$-N$_{\rm U}^{\rm (p)}$ transition (see 
\cite{deGennes1} for a detailed discussion on this point).

The manner in which the first-order N$_{\rm U}^{\rm (r)}$-N$_{\rm U}^{\rm (p)}$ transition connects to the biaxial nematic
phase is also peculiar. As the packing fraction is increased following the 
N$_{\rm U}^{\rm (r)}$--N$_{\rm U}^{\rm (p)}$ binodals, first the N$_{\rm U}^{\rm (r)}$ changes to
N$_{\rm B}$ at a continuous transition, while the transition gap decreases and eventually vanishes, see Fig. 
\ref{fig3}(b). Then the transition reappears as a first-order transition and again becomes continuous at a tricritical 
point.  These phenomena occur in a very small region of the phase diagram.

The spinodal curves in the case $\kappa_2=5$ present some important changes with respect to the previous case.
One is that, due to the higher aspect ratio, the PS phase is very unstable. For high $\kappa_1$ there appear
unstable fluctuations in the N$_{\rm U}^{\rm (p)}$ (not the I) phase, and these fluctuations are now of columnar
symmetry, C$_{\rm U}^{\rm (p)}$. This symmetry involves particles arranged with their shortest axes parallel to the 
columns. As the aspect ratio increases from unity, columns become unstable, and the spinodal changes to 
S$_{\rm U}^{\rm (p)}$. For still larger $\kappa_1$ this spinodal meets the biaxial nematic region, and reappears
in this region as a S$_{\rm B}^{\rm (p)}$ spinodal. This phase consists of particles arranged in layers with
their shortest axes perpendicular to the layers, and with their long axes oriented in the layer planes.
Again the spinodal changes its nature and joins the spinodal curve of the S$_{\rm B}^{\rm (r)}$ phase; 
this is a biaxial smectic phase
with the long particles axes perpendicular to the layers and with the secondary particle axes oriented in
the layer planes. Therefore, as the aspect ratio $\kappa_2$ increases, the smectic phase tends to develop
two types of symmetries, namely rod- and plate-like symmetries. These phases were not observed in \cite{Taylor}
due to the low aspect ratio chosen.

\begin{figure}[h]
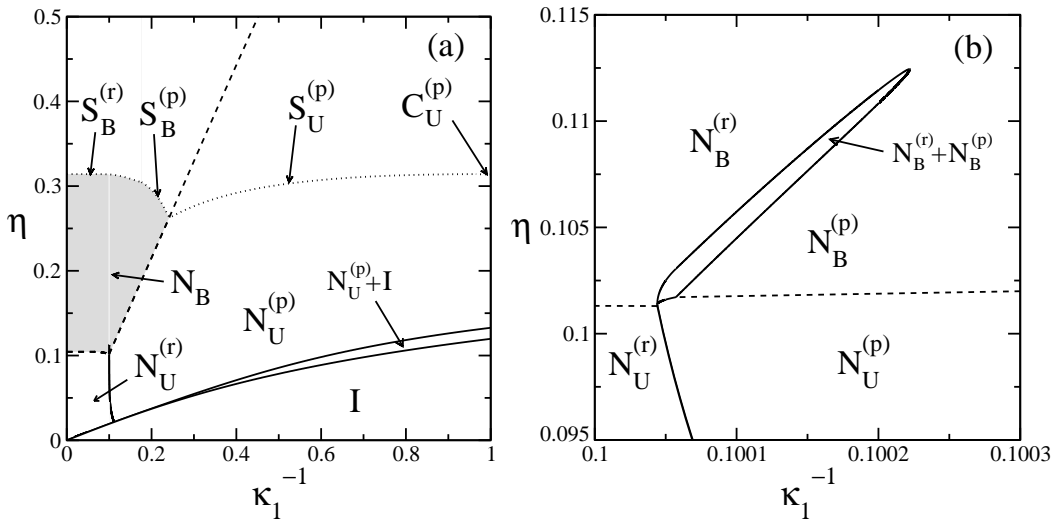

\includegraphics[width=2.6in]{Fig4a.eps}
\includegraphics[width=2.85in]{Fig4b.eps}
\caption{(a) Phase diagram in the plane $\eta-\kappa_1$
for hard biaxial parallelepipeds with $\kappa_2=10$. (b) A zoom of the
region about $\kappa_1^{-1}\approx 0.1$. Labels as in Fig. \ref{fig2}.}
\label{fig5}
\end{figure}

Finally, we note that the region where the N$_{\rm B}$ phase is stable has considerably increased with respect to
the $\kappa_2=2.5$. Nevertheless, we must bear in mind that the nematic-to-smectic transitions could be of
first order, and therefore that the stability region of the N$_{\rm B}$ phase predicted in Fig. \ref{fig3}(a)
could be much reduced or even vanish. Also, the spinodal lines shown, which bound the N$_{\rm B}$ region from
above, have been calculated from the stable phases at the particular density chosen. But free-energy 
differences between unstable uniaxial and stable biaxial nematic phases are very small, as can be seen in Fig. \ref{frees}(a).
In fact, the unstable N$_{\rm U}^{\rm (r,p)}$ phases bifurcate 
[filled triangle and square in Fig. \ref{frees}(a)] to a non-uniform phase before the stable N$_{\rm B}$ phase
does (filled circle), i.e. inside the shaded region labelled N$_{\rm B}$ in Fig. \ref{frees}(a). It could 
happen that the branch to which these unstable N$_{\rm U}^{\rm (r,p)}$ phases bifurcate become more
stable than the N$_{\rm B}$ phase before its corresponding bifurcation point. This scenario
would reduce (but not completely eliminate) the shaded region. For the sake of completeness and illustration, the 
behaviour of the order parameters along the stable free-energy branch in plotted in Fig. \ref{frees}(b);
in particular, we see why $\Delta$ is a more convenient parameter than $\delta$ to represent biaxial
order in this system. Note the discontinuity in $Q$ indicated by the symbols, caused by the
first-order nature of the N$_{\rm U}^{\rm (p)}$-N$_{\rm U}^{\rm (r)}$ transition.

Our final result concerns the case $\kappa_2=10$, Fig. \ref{fig5}. 
Here the N$_{\rm U}^{\rm (r)}$-N$_{\rm U}^{\rm (p)}$ transition remains of first order and again 
ends in a Landau point as it joins the I-N$_{\rm U}^{(r,p)}$
transition. The transition gap is considerably decreased, and this is probably related 
to the discretization of particle orientations, which promotes strong orientational
ordering of particles with high aspect ratios. But an important new feature is that now 
there appear two biaxial nematic phases, N$_{\rm B}^{\rm (r)}$ and N$_{\rm B}^{\rm (p)}$, coexisting
in a narrow density range [see Fig. \ref{fig5}(b)].
The spinodal curves appear at more or less similar values of the packing fraction, but the
nematic stability begins at much lower values, with the consequence that
the predicted stability region of the biaxial nematic phase is much increased.

\section{Conclusions}
\label{IV}

Since the motivation of the present work was the observation of biaxial nematic phases in goethite \cite{Vroege2},
it is natural to ask if our results reproduce the experimental findings and if our theory could
therefore be used to predict phase behaviour in these systems with any confidence. 
In the experiments reported in \cite{Vroege3}, which summarise the work done on goethite by the Utrecht group
up to now, $\kappa_2$ is in the range $2.2-3.0$ (probably a bit lower if one takes into account the
Debye screening lengths of the particles in the solution) and $6.2\agt\kappa_1\agt 3.1$.
Only in the case $\kappa_1\approx\kappa_2\approx 3$ is a biaxial nematic phase found. This ratio
corresponds to a phase diagram in between those of Figs. \ref{fig2} and \ref{fig3} but,
according to our calculations, it would not correspond to favourable particle size ratios
to easily observe the N$_{\rm B}$ phase. Our prediction is that
higher values of both $\kappa_1$ and $\kappa_2$ (with $\kappa_1\agt\kappa_2$) 
should be explored to find stable biaxial nematic phases
in a wider concentration range. However, two aspects require further consideration in order to
obtain more reliable predictions. One is the calculation of absolute free-energy minima of the
density functional for the non-uniform phases, which could give rise to shifted stability regions in case first-order 
transitions were found. The other is the effect of particle size polydispersity (which in the
experiments is in the range $0.25-0.55$). This effect may be very important, as polydispersity tends to
preclude the occurrence of phases with partial positional order (leaving wider N$_{\rm B}$
stability windows) and induce demixing phenomena.
These considerations point to the need to incorporate particle size polydispersity in the calculations
if one is to establish closer contact with the experiments. However, from the computational point of view,
this task is exceedingly difficult, since there exist three, in principle
not completely independent, polydispersity parameters \cite{vandenPol}. Efforts in this
direction are now being undertaken in our group. For the moment, 
we hope that our present work represents a step forward in understanding the phase behaviour of biaxial 
particles and can help foster the development of FMT studies in the direction of more complex particle shapes
\cite{Schmidt,FMT-PRL}.

\acknowledgements
The authors gratefully acknowledge the support from the Direcci\'on General de Investigaci\'on
Cient\'{\i}fica y T\'ecnica under Grants Nos. FIS2010-22047-C05-01 and FIS2010-22047-C05-04, and from the
Direcci\'on General de Universidades e Investigaci\'on de la Comunidad
de Madrid under Grant No. S2009/ESP-1691 and Program MODELICO-CM.


\begin{references}
\bibitem{C} This is because the values of the order parameters change smoothly from those typical
of a S$_{\rm B}$ phase to those typical of a C$_{\rm U}$ phase.
\end{references}
\end{document}